\documentclass[twocolumn,prl,superscriptaddress,floatfix,aps]{revtex4}
\newcommand{\be}{\begin{equation}}
\newcommand{\ee}{\end{equation}}

\usepackage{graphicx}
\usepackage{dcolumn}
\usepackage{bm}
\usepackage{psfig}
\usepackage{longtable}

\begin{document}
\title{Coupled Electron Ion Monte Carlo Calculations of Dense Metallic
Hydrogen}
\author{Carlo Pierleoni}
\affiliation{INFM and Department of Physics, University of
L'Aquila, Via Vetoio, I-67010 L'Aquila, Italy}
\author{David M. Ceperley}
\affiliation{University of Illinois at Urbana-Champaign, Urbana,
IL 61801, USA}
\author{Markus Holzmann}
\affiliation{LPTL, UMR 7600 of CNRS, Universit\'e P. et M. Curie, Paris, France}


\begin{abstract}
We present a new Monte Carlo method which couples Path Integral
for finite temperature protons with Quantum Monte Carlo for ground
state electrons, and we apply it to metallic hydrogen for
pressures beyond molecular dissociation. We report data for the
equation of state for temperatures across the melting of the
proton crystal. Our data exhibit more structure and higher melting
temperatures of the proton crystal than Car-Parrinello Molecular
Dynamics results. This method fills the gap between high
temperature electron-proton Path Integral and ground state
Diffusion Monte Carlo methods.
\end{abstract}

\maketitle



The knowledge of the physical
properties of hydrogen in a wide range of thermodynamic conditions is a key
problem in planetary and high pressure physics\cite{hubbard84,MaoHemley94}.
In the search for the
metallization transition three different insulating  molecular
crystal phases have been clearly observed so far in diamond anvil
cell experiments up to 3.2Mbar\cite{loubeyre} at room temperature
and below. Metallization has been obtained in shock wave
experiments for a warm dense molecular liquid\cite{nellis} but
properties at finite temperature and/or at higher pressure are
largely unknown because experiments are increasingly difficult.

A large body of theoretical investigations of high pressure hydrogen have
appeared over the years\cite{maksimivshilov}.
They helped the understanding of the experimental observations and
hold out the prospect of predicting the room temperature
metallization pressure and the phase diagram at higher pressure.
However the present understanding of high pressure hydrogen is
unsatisfactory because
1) energy differences among different crystalline phases are small
requiring a very accurate total energy method to determine the
stable crystalline phase and locate transition lines; 2) size
effects are large in metallic and quasi-metallic systems and
Brillouin zone sampling is extremely important for accurate total
energy calculations; 3) proton quantum effects are important 
and can influence the energetic ordering of crystal phases
; 4) an accurate theoretical prediction of
metallization may require accuracy beyond that of the
LDA+GGA Density Functional Theory \cite{stadelemartin,johnsonaschroft}.

Here we describe a method based on Quantum Monte Carlo (QMC)
calculation of the electronic energy for quantum mechanical
protons able to sample efficiently the protonic configurational
space and spontaneously find the stable phase of the system within
the Born-Oppenheimer approximation. Previous QMC studies of
hydrogen at $T=0$ have treated electrons and protons at the same
level of description and become inefficient in following the
evolution of particles of very dissimilar mass ($m_p/m_e=1836)$.
Moreover, the interesting effects of temperature are absent in
this procedure. Nonetheless, they have established that pressure
dissociation of hydrogen molecules at T=0K occurs at
$r_s=[3/(4\pi~n_e)]^{1/3}=1.31~(P\sim 3Mbars)$\cite{dmc81}, where
$n_e$ is the electronic number density. Upon dissociation the
molecular crystal transforms to a proton lattice of diamond
structure and later to a lattice of cubic symmetry (bcc) at
$P\ge8Mbars$\cite{natoli93,natoli95}. At finite temperature
Restricted Path Integral Monte Carlo (RPIMC)\cite{como95} has been
used to predict the equation of state (EOS) and to investigate the
occurrence of the plasma phase transition\cite{pimc}. In RPIMC,
both electrons and protons are at finite temperature but it is
efficient only for temperatures above 1/20 of the electronic Fermi
temperature (roughly $3 \times 10^4K$ at $r_s=1$).
The new method described here, called Coupled Electronic-Ionic
Monte Carlo (CEIMC)\cite{dewing01,dmc03}, is able to fill the gap
between the RPIMC and the ground state QMC methods.
We study metallic hydrogen in a range of densities and
temperatures where molecules are absent and where protons undergo
a solid-fluid transition. We report results for the EOS and give a
qualitative location of the transition line. 

In the CEIMC method the proton degrees of freedom are advanced by
a Metropolis algorithm in which the energy difference between the
actual state $S$ and the trial state $S'$ is computed by a Quantum
Monte Carlo calculation (either variational (VMC) or reptation (RQMC)
\cite{BaroniMoroni}).
The energy difference is 
affected by statistical noise which would bias the MC sampling.
Unbiased sampling of the proton configurations can be achieved by
the penalty method\cite{penalty}, a generalization of the
Metropolis algorithm.

We sample the electronic degrees of freedom according to the sum
of the electronic distribution functions ({\it e. g.} the square
of the trial wave function in VMC) for the $S$ and $S'$ states,
and we compute the energies for the two states as correlated
sampling averages\cite{dewing01,dmc03}, thereby reducing the
noise.
Analytic trial wave functions including backflow and three-body
correlation\cite{hcpe03} have been used in most of our
calculations. These functions are particularly appropriate to our
methods since: 1) they are quite accurate; 2) they are free of
adjustable parameters so do not require optimization; 3) their
computational cost is much less then solving the Kohn-Sham
equations as was done in previous QMC
calculations\cite{natoli93,natoli95}, in particular for a random
arrangement of several tens of protons.

To go beyond VMC,
we implemented a Reptation
Quantum Monte Carlo algorithm (RQMC)\cite{BaroniMoroni} to sample
more accurately the electronic ground state.
Similar to Diffusion Monte Carlo (DMC),
RQMC projects the trial wavefunction to the ground state
within the Fixed-Node approximation.
The high quality of our trial wave functions makes it possible to
relax to the ground state with a very limited number of time
slices. In RQMC the electronic path space is sampled by a
reptation algorithm in which, at each step, a new link is added to
one end of the path and an existing link is deleted from the other
end, subject to a ``Metropolis'' acceptance/rejection step.
To speed up convergence we have introduced the ``bounce'' algorithm
in which
the growth direction is reversed only when a move is rejected\cite{CEIMC-new}.
RQMC is particularly suited for computing energy differences.

To reduce finite size effects in
metallic systems, we average over
twisted boundary conditions (TABC) when computing electronic energies
({\it i.e.} we integrate over the Brillouin zone
of the super cell)\cite{lzc01,dmc03}.
All properties are averaged
over 1000 different k-points.
For the typical protonic displacement, we
compute the energy difference over
100 electronic steps/k-point.
After averaging over k-points, the noise level is
small enough to simulate temperatures as low as 100K\cite{dmc03}.

We represent protons by imaginary time path integrals without
considering the statistics of the protons.
(those effects are negligible in this temperature-density range.)
For efficiency, it is important
to minimize the number of protonic time slices. We have used the
pair action of an effective proton-proton potential and treated
the difference between the true Born-Oppenheimer energy and the
effective potential with the primitive approximation\cite{rmp95}.
With this action, we find that a proton imaginary time step
$\tau_p=0.3\times10^{-3} K^{-1}$ is appropriate for $r_s\ge 1$ so
that few tens of time slices allow for calculations above 100K.
When coupled with TABC, we can, at each proton move, randomly
assign a subset of k-points to each protonic slice without
introducing a detectable systematic effect. This
strategy allows one to simulate quantum protons 
at essentially the
same computational cost as classical protons
, except for the
slower relaxation of the protonic paths.

\begin{figure}
\centerline{\psfig{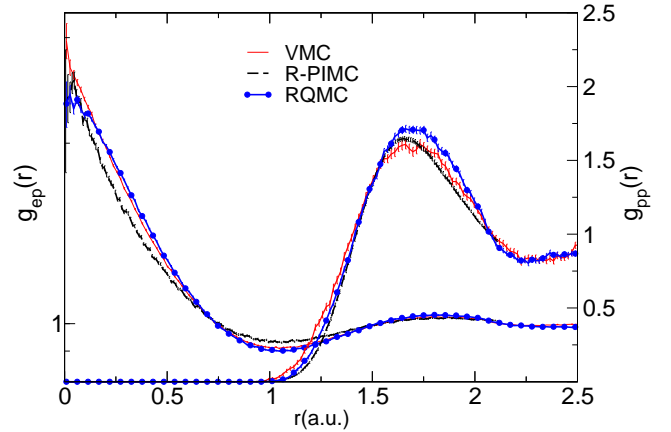}}
\caption{CEIMC-RPIMC comparison for electron-proton and proton-proton
correlation function at $r_s=1, T=5000K, N_p=N_e=16$.}
\label{fig:pimc}
\end{figure}
In order to assess the accuracy of the CEIMC method we first
consider a system of $N_p=N_e=16$ at $r_s=1$ and T=$5000K$ and
compare with RPIMC.
The CEIMC-RQMC calculation, performed with
$\tau_e=0.0125H^{-1},\beta_e=0.5H^{-1}$ (41 electronic time
slices), provides a total energy lower than the VMC estimate by
$4(2)mH/atom=1260(630)K$.
However, the VMC and RQMC pressures agree within error bars.
Comparison between VMC and RQMC pair correlation functions is also
very good (see figure~\ref{fig:pimc}). The VMC and RQMC
$g_{ep}(r)$'s are superimposed except at distance below $0.2 a_0$
($a_0=0.529 \AA$ is the Bohr radius), due to time step error in
RQMC. As for $g_{pp}(r)$, RQMC curve is slightly more structured
than the VMC one. In RPIMC, such ``low'' temperatures (the Fermi
temperature at $r_s=1$ is $1.84 H=5.8~10^5K$) can be reached only
by imposing less realistic ground state nodal
restriction\cite{como95,dmc03}. RPIMC data, obtained with free
particle nodes and 1000 time slices, agrees with CEIMC ones. CEIMC
computed $g_{ep}(r)$ exhibits slightly more structure than RPIMC,
and since thermal effects on the electrons should be largely
negligible in such conditions, we attribute the observed
difference to the more accurate nodal structure of CEIMC compared
to RPIMC.

\begin{figure}
\centerline{\psfig{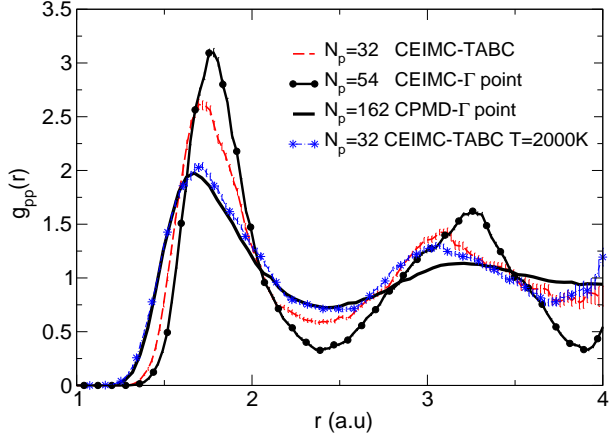}}
\caption{Pair correlation function at $r_s=1, T=1000K$. Comparison
between CEIMC-VMC-TABC with $N_p=32$, CEIMC-VMC-PBC $N_p= 54$ s
and CPMD-LDA $N_p=162$ (simulation with $N_p=54$ is identical).
Data from CEIMC-VMC-TABC at T=2000K (stars) are also reported. }
\label{fig:lda}
\end{figure}

Next we compare with Car-Parrinello Molecular Dynamics (CPMD)
simulation\cite{kh95} which uses the LDA computed forces.
Figure \ref{fig:lda} shows
that CEIMC-VMC $g_{pp}(r)$'s for classical protons exhibit considerably
more structure than does LDA. 
CPMD simulations considered systems of classical protons with a
closed shell (in reciprocal space), and only the $\Gamma$ point.
We compare with two different CEIMC calculations for classical
protons, namely an open shell system ($N_p=32$) with the TABC, and
a closed shell system ($N_p=54$) with the $\Gamma$ point only. For
the latter case, we find that the $g_{pp}(r)$ from VMC and RQMC
(not shown) agree; but they exhibit more structure than CPMD. The
TABC one is in the liquid state, while the simulation using only
the $\Gamma$ point, initially prepared in a liquid state from
temperature quenching, exhibits the onset of spontaneous
crystallization. The larger correlation in CEIMC with respect to
CPMD is compatible with our early estimate of the melting
temperature of the fcc crystal of classical proton between 1000K
and 1500K \cite{dmc03} at variance with the the LDA estimate of
350K (for the bcc crystal)\cite{kh95}.
The observed discrepancy between CEIMC and CPMD is surprising
since LDA is generally believed to be accurate at high density.
However a previous study of hydrogen at $r_s=1.31$\cite{natoli93}
reported that differences in energy among several crystal
structures obtained within LDA are smaller than energy differences
from Diffusion MC by roughly a factor of two. Also zero point
energies in QMC were roughly twice the LDA estimates (from the
harmonic approximation). This suggests that the Born-Oppenheimer
surface from LDA is flatter than the more accurate one from QMC.
Moreover there is a known issue in computing the ionic temperature
in CPMD;
the simple estimate based on the ionic kinetic energy provides
only a lower bound for the true
temperature\cite{tangneyscandolo02}. Tracing the origin of the
observed discrepancy between CEIMC and CPMD results would deserve
an independent study. Here we just note that better agreement is
observed between CPMD results at temperature T and CEIMC data at
temperature $2T$ for $300 \leq T \leq 3000$, see for instance
figure \ref{fig:lda}.

RQMC is roughly an order of magnitude more expensive than VMC.
Therefore, it is important to establish the accuracy of VMC before
performing a systematic study of the equation of state.
In table \ref{tab:table1} we compare VMC and RQMC 
at $T=5000 K$ and $r_s=1.2$, the lowest density we have
considered. Calculations are done at the $\Gamma$ point and a
projection time of $\beta_e=0.68 H^{-1}$ for RQMC.
We have checked that for protons in bcc lattice, the energy has
converged to its ground state value for $\beta\sim 0.6$.
We find that the VMC energy is systematically
higher by roughly $7.6(2)mH/at=2400(60)K$,
while the VMC pressure is systematically lower by $0.03(1) Mbars$.
The error on the energy is expected to be independent of the
temperature and to decrease with increasing density. Even though
the amount of energy missing in VMC is quite large on the proton
energy scale, we only observe a minor effect on $g_{pp}(r)$;
energy differences are quite accurate within VMC.
\begin{table}
\caption{Comparison between VMC and RQMC energies and pressures
for  $r_s=1.2, T=5000K, N_p=N_e=54$, PBC. $\tau=0 $ is the
estimate with RQMC time step errors removed by extrapolation.
$\sigma^2$ is the variance of the local energy in VMC.}
\label{tab:table1}
\begin{tabular}{|c|ccccc|}
\hline
$\tau_e   $&E$_{tot}$(h/at) & $\sigma^2$ & E$_{kin}$  & E$_{pot}$  & P (Mbars)
\\\hline
vmc       &  -0.4694(2)  & 0.0472(4)  & 0.8812(4)&-1.3508(4)& 5.55(1)   \\
0.01      &  -0.4768(4)  &   -----    & 0.8850(6)&-1.3618(6)& 5.50(1)   \\
0.00      &  -0.47696    &   -----    & 0.89112  &-1.36808  & 5.581     \\
\hline
\end{tabular}
\end{table}
On the basis of the above results, we performed a systematic study
of the Equation of State (EOS) using VMC.
In table \ref{tab:table2} we report total, kinetic and potential
energies, pressure, the Lindemann ratio for bcc crystal, and the
proton kinetic energy. The latter quantity can be compared to
$3K_BT/2$ (last column). The zero point proton motion affects not
only the proton kinetic energy but also increases the electronic
kinetic energy and, to a smaller extent, the configurational
energy. At $r_s=1$ and $T=500K$ we find a total energy increase of
$14.9(2)mH/at=4670(60)K$ of which $2020(30)K$ comes from the
proton kinetic energy, $2200(20)K$ the electronic kinetic energy,
and $450(10)K$ the configurational energy. Residual finite size
effects have been estimated from static lattice calculation at
$r_s=1$ to be of the order of $10mH/at$ on the energy, and
$0.21~Mbars$ on the pressure. The transition line, estimated by
the dynamical observation of melting, is located between 1000K and
2000K at $r_s=0.8$, between 500K and 1000K at $r_s=1.0$ and close
to 1000K at $r_s=1.2$. Indeed at the latter density and at T=1000K
the system is able to sustain both liquid and crystal states for
the entire length of our simulations (80000 protonic steps).

\begin{table*}
\caption{Energy and pressure for a system of $N_p=54$ quantum
protons with VMC-TABC. Units of energy are hartrees/proton.
$M_p$ is the number of protonic time slices ($M_p=1$ means classical protons).
$\gamma$ is the rms deviation divided
by the nearest neighbor distance for a bcc lattice.} \label{tab:table2}
\begin{tabular}{|c|c|c|ccc|c|c|cc|} \hline

 $r_s$&T(KK)&$M_p$& $E$& $E_{kin}$   &  $E_{pot}$ &$P(Mbars)$&$\gamma_L$ & $K_p \times 10^2$ & $K_p^{cl}\times 10^2$ \\
 \hline
  0.8 & 0.5 &  16 & -0.0594(2) & 1.8419(1) & -1.9033(1) & 81.07(3) & 0.169(1)  & 1.57(3)&  0.2375\\
      & 1.0 &  16 & -0.0586(4) & 1.8428(4) & -1.9034(1) & 81.16(3) & 0.183(1)  & 1.53(4)&  0.475 \\
      & 2.0 &   8 & -0.0522(4) & 1.8338(4) & -1.9018(1) & 81.69(3) &   ---     & 1.78(3)&  0.950 \\
      & 3.0 &   4 & -0.0442(4) & 1.8538(6) & -1.9000(2) & 82.33(6) &   ---     & 2.14(7)&  1.425 \\
      & 4.0 &   4 & -0.0382(8) & 1.8590(8) & -1.8991(1) & 82.83(6) &   ---     & 2.57(7)&  1.900 \\
      & 6.0 &   2 & -0.0268(8) & 1.8688(8) & -1.8974(2) & 83.80(6) &   ---     & 3.29(4)&  2.850 \\
      &10.0 &   1 &  0.016(1)  & 1.8886(8) & -1.8934(4) & 85.78(9) &   ---     & 4.750  &  4.750 \\
\hline
  1.0 & 0.5 &   8 & -0.3512(2) & 1.2142(2) & -1.5655(1) & 20.101(3)& 0.177(1)  & 0.97(2)&  0.2375 \\
      & 1.0 &   4 & -0.3480(2) & 1.2176(2) & -1.5657(1) & 19.68(1) &   ---     & 1.07(2)&  0.475  \\
      & 2.0 &   4 & -0.3430(2) & 1.2260(4) & -1.5653(1) & 20.65(1) &   ---     & 1.44(2)&  0.950 \\
      & 3.0 &   2 & -0.3356(4) & 1.2298(4) & -1.5655(1) & 20.83(1) &   ---     & 1.72(3)&  1.425 \\
      & 5.0 &   1 & -0.3262(6) & 1.2390(6) & -1.5652(1) & 21.26(2) &   ---     & 2.375  &  2.375 \\
      &10.0 &   1 & -0.2888(6) & 1.2740(4) & -1.5630(2) & 22.95(3) &   ---     & 4.750  &  4.750 \\
\hline
  1.2 & 0.3 &  10 & -0.46610(4)& 0.8776(1) & -1.3437(1) & 5.554(1) & 0.134(1)  & 0.59(1)&  0.1425 \\
      & 0.5 &   8 & -0.4661(1) & 0.8792(1) & -1.3439(1) & 5.594(3) & 0.177(2)  & 0.67(1)&  0.2375 \\
      & 1.0 &   4 & -0.4632(1) & 0.8811(2) & -1.3443(2) & 5.641(3) & 0.196(3)  & 0.77(1)&  0.475 \\
      & 1.0 &   4 & -0.4610(2) & 0.8858(2) & -1.3468(1) & 5.735(6) &  liquid   & 0.77(1)&  0.475 \\
      & 2.0 &   4 & -0.4552(2) & 0.8918(2) & -1.3469(1) & 5.893(6) &   ---     & 1.19(3)&  0.950 \\
      & 3.0 &   2 & -0.4492(4) & 0.8996(3) & -1.3488(1) & 6.08(2)  &   ---     & 1.53(3)&  1.425 \\
      & 5.0 &   1 & -0.4386(6) & 0.9106(4) & -1.3492(2) & 6.37(2)  &   ---     & 2.375  &  2.375 \\
      &10.0 &   1 & -0.4036(6) & 0.9478(4) & -1.3514(1) & 7.34(2)  &   ---     & 4.750  &  4.750 \\
\hline
\end{tabular}
\end{table*}
In conclusion we have developed a new and efficient Quantum Monte
Carlo Method to study low temperature quantum protons and ground
state electrons which is a major improvement over previous QMC and
DFT-LDA based methods. It allows for simulations of many-body
hydrogen using QMC for the electronic energies. We have developed
efficient procedures to include protonic path integrals and
k-point sampling. We have applied it to metallic hydrogen beyond
molecular dissociation and investigated the solid-fluid transition
of the protons. The present methodology can be extended in several
ways. Constant pressure algorithm would be useful to study
structural phase transitions. However for metallic systems, we
have found that level crossings, arising from changes in the shape
of the simulation box, considerably increase the noise level and
makes our correlated sampling procedure inefficient. The method
can be easily extended to the insulating molecular phase by
replacing the metallic trial functions with localized molecular
orbitals\cite{dewing01,dmc03}. A study of the melting line of
molecular hydrogen is in progress. Consideration of the
metal-insulator transition requires a trial function that goes
smoothly from metallic to localized orbitals. We are investigating
an accurate and efficient form for this. Extension of the present
method to more complex elements is straightforward, provided we
have efficient trial functions.

Early aspects of the CEIMC algorithm were developed in
collaboration with M. Dewing. We have the pleasure to thank
J.P.Hansen and J.Kohanoff for useful discussions and for providing
their CPMD data, and S.Scandolo and R.M. Martin for illuminating discussions.
This work has been supported by a visiting grant from INFM-SezG and by
MIUR-COFIN-2003. Computer time has been provided by NCSA
(Illinois), PSC (Pennsylvania) and CINECA (Italy) through the INFM
Parallel Computing initiative.


\end{document}